\documentclass[twocolumn,superscriptaddress,amsmath,amssymb,aps,pra]{revtex4-1}

\usepackage{graphicx}
\usepackage{dcolumn}
\usepackage{bm}

\usepackage{comment}

\usepackage[colorlinks,urlcolor=blue,citecolor=blue,linkcolor=blue]{hyperref}

\renewcommand{\k}{{\bf k}}
\newcommand{\p}{{\bf p}}

\newcommand{\q}{{\bf q}}

\renewcommand{\u}{{\bf u}}
\renewcommand{\v}{{\bf v}}

\newcommand{\bra}[1]{\left\langle{#1}\right|}
\newcommand{\ket}[1]{\left|{#1}\right>}

\newcommand{\eb}{E_{\rm B}}
\newcommand{\eq}{\epsilon_{\q}}
\newcommand{\ek}{\epsilon_{\k}}

\newcommand{\ekone}{\epsilon_{\k_1}}
\newcommand{\ektwo}{\epsilon_{\k_2}}

\newcommand{\nn}{\nonumber}
\newcommand{\beq}{\begin{equation}}
\newcommand{\eeq}{\end{equation}}
\newcommand{\inv}{^{-1}}

\newcommand{\binomial}[2]{\left(\begin{array}{c}#1\\#2\end{array}\right)}
\newcommand{\sch}{Schr{\"o}dinger }

\newcommand{\kb}{\kappa_\mathrm{B}}

\newcommand{\sout}[1]{}

\begin{document}

\title{Few-body states of bosons interacting with a heavy quantum impurity}

\author{Shuhei M.\ Yoshida}
\thanks{These authors contributed equally to this work.}
\affiliation{Department of Physics, The University of Tokyo, Tokyo 113-0033, Japan}
\affiliation{School of Physics and Astronomy, Monash University, Victoria 3800, Australia}

\author{Zhe-Yu Shi}
\thanks{These authors contributed equally to this work.}
\affiliation{School of Physics and Astronomy, Monash University, Victoria 3800, Australia}

\author{Jesper Levinsen}
\affiliation{School of Physics and Astronomy, Monash University, Victoria 3800, Australia}

\author{Meera M.\ Parish}
\affiliation{School of Physics and Astronomy, Monash University, Victoria 3800, Australia}

\date{\today}

\begin{abstract}
We consider the problem of a fixed impurity coupled to a small number $N$ of non-interacting bosons. 
We focus on impurity-boson interactions that are mediated by a closed-channel molecule, as is the case for tuneable interatomic interactions in cold-atom experiments. We show that this two-channel model can be mapped to a boson model with effective boson-boson repulsion, which enables us to solve the three-body ($N=2$) problem analytically and determine the trimer energy for impurity-boson scattering lengths $a>0$. By analysing the atom-dimer scattering amplitude, we find a critical scattering length $a^*$ at which the atom-dimer scattering length diverges and the trimer merges into the dimer continuum. 
We furthermore calculate the tetramer energy exactly for $a>0$ and show that the tetramer also merges with the continuum at $a^*$.
Indeed, since the critical point $a^*$ formally resembles the unitary point $1/a = 0$, we find that all higher-body bound states (involving the impurity and $N>1$ bosons)
emerge and disappear at both of these points.
We show that the behavior at these ``multi-body resonances'' is universal, since it occurs for any model with an effective three-body repulsion involving the impurity. Thus, we see that the fixed-impurity problem is strongly affected by a three-body parameter even in the absence of the Efimov effect.
\end{abstract}


\maketitle

\section{Introduction}

\begin{figure}[th]
    \centering
    \includegraphics[width=0.96\columnwidth]{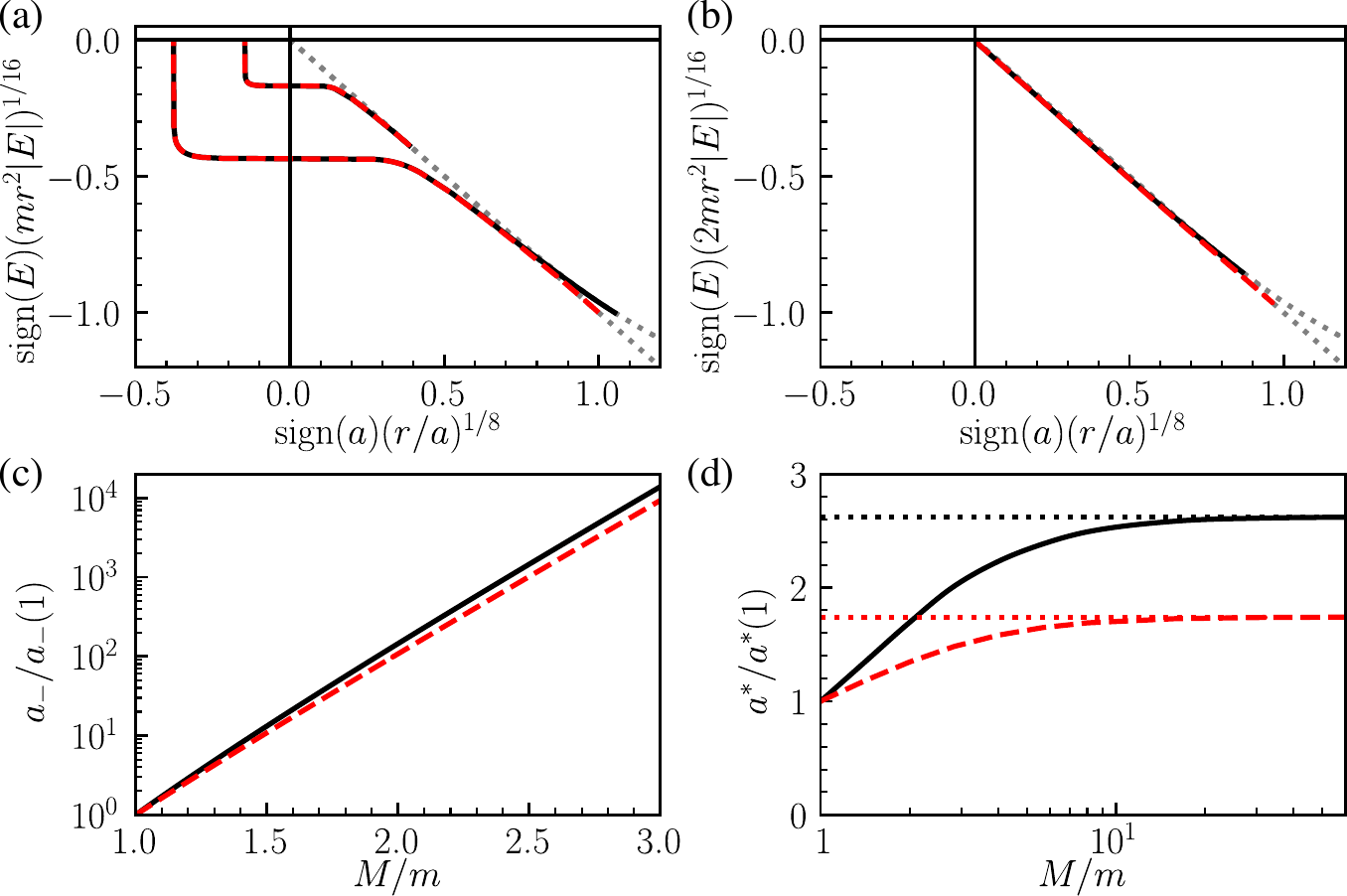}
    \caption{(a,b) Trimer spectrum of two identical bosons of mass $m$
      interacting near-resonantly with an impurity atom of mass (a)
      $M=m$ and (b) $M=\infty$. The
      spectrum in (a) was calculated in Ref.~\cite{Yoshida2018} and
      shows how trimers can be Borromean 
      when $M$ is finite.
      The factor of 2 in the vertical axis label in (b) ensures that the dimer lines (dotted) are identical to those in (a).
      (c) Three-body parameter $a_-$ and (d) atom-dimer resonance $a^*$ in units of their value at $M=m$, both as
      a function of mass ratio. The results are obtained within
      the two-channel model~\eqref{eq:2ch} with an effective range
      $r_0$ (black, solid),
      or within the $\Lambda$-model (red, dashed). In (a,b) we take $r=|r_0|=0.5487/\Lambda$.}
    \label{fig:mass}
\end{figure}

The vigorous investigation of ultracold Bose gases with
near-resonant two-body interactions has to a large extent been driven
by the peculiar nature of the associated few-body spectrum. 
Most notably,
in 1970 Efimov predicted~\cite{Efimov1971} that a system of three
identical bosons with short-range interactions features an infinite number of three-body 
bound states in the limit where the two-body scattering
length $a\to\infty$. These so-called Efimov trimers can exist even in
the absence of a two-body bound state --- a property referred to as Borromean --- and they form a geometric
spectrum that satisfies a discrete scaling symmetry~\cite{Efimov1970,Braaten2006}: each
trimer can be related to the others via a discrete rescaling of the scattering
length and energy, i.e., $a\to \lambda^na$ and $E\to \lambda^{-2n}E$, with
$n$ integer and $\lambda$ a scaling factor, as illustrated in
Fig.~\ref{fig:mass}(a). 

While Efimov's original prediction was in the context of nuclear
physics~\cite{Efimov1971}, the first experimental evidence of Efimov physics was found
in 2006 in an ultracold atomic Bose gas~\cite{Kraemer2006}. Indeed, 
cold atoms offer the ideal test bed for the investigation of
few-body physics due to
the tunability of the two-body interactions using Fano-Feshbach
resonances~\cite{Chin2010}, and this has recently allowed the
experimental observation of the second trimer in Efimov's
scenario~\cite{Huang2014}. The rich few-body spectrum extends beyond
trimers to larger clusters, such as tetramers~\cite{Stecher2009,Ferlaino2009} and even pentamers and larger bound
clusters~\cite{Stecher2011PRL,Blume2014}. These developments have stimulated much recent theoretical and experimental progress, as reviewed in Refs.~\cite{naidon2017,greene2017,dIncao2017}.

The scaling factor $\lambda$ depends sensitively on the scenario under
consideration. For instance, in a system of identical bosons, we have
$\lambda\simeq22.7$~\cite{Efimov1970}, while in a mass-imbalanced system of a single
particle with mass $M$ resonantly interacting with two particles of
mass $m$, $\lambda$ is greatly reduced when
$M\ll m$~\cite{Efimov1973}. Such a situation is realized in Cs-Li mixtures, which thus allowed the observation of several Efimov trimers~\cite{Tung2014,Ulmanis2016}.
The opposite limit of a heavy ``impurity'' atom has received comparatively
less attention since this is Efimov unfavored, with $\lambda\to\infty$
when $M/m\to\infty$. As a consequence, as shown in
panels~\ref{fig:mass}(b,c), the scattering length $a_-$ at which the
ground-state Efimov trimer crosses into the three-atom continuum
becomes infinite, and trimers cease to be Borromean. The parameter
$a_-$ is of central importance in Efimov physics 
since it corresponds to an additional parameter (the three-body parameter) that
sets the scale of the ground state in the Efimov
spectrum~\cite{Berninger2011,Wang2012a,naidon2014b} and prevents the system from collapsing~\cite{Thomas1935} in the so-called fall to the centre~\cite{LL}. 
The divergence of $a_-$ thus suggests that the three-body parameter drops out of the problem.

In this paper, which accompanies Ref.~\cite{shortpaper}, we consider the scenario of an infinitely heavy impurity atom with $N$ non-interacting identical bosons. We argue that --- even in the absence of Efimov physics --- the system is pathological without a high-energy cutoff on the three-body dynamics. 
Specifically, if the impurity corresponds to a static potential with a bound state of energy $-\eb$, then the ground-state energy of the system would simply be $E=-N\eb$ in the absence of a cutoff akin to the three-body parameter. 
This implies that the energy becomes arbitrarily large with increasing $N$, such that the details of the physics at arbitrarily short length scales will become relevant.

Here we show that an effective three-body repulsion (involving two bosons and the impurity) leads to a significant renormalization of the ground-state energy, from $E=-N\eb$ to
\begin{align}
        E\simeq
        - N\eb
        + 
        \frac{N(N-1)\pi}{\log a}\eb,
\end{align}
in the limit when $1/a\to0^+$.
Indeed, this repulsion  leads to a surprisingly strong renormalization of the ground state already when $N=2$.
The logarithmic correction is negligible only in an exponentially small region close to unitarity, and it is universal in the sense that it does not depend on the manner in which an effective three-body repulsion is introduced. We thus find that clusters can form between the impurity and an arbitrary number of bosons, and that they all merge into the scattering continuum when $1/a\to0^+$. Due to this property, we term the unitary point a ``multi-body resonance''~\cite{shortpaper}.

Our results are based on two models which both feature an effective three-body repulsion involving the impurity.
Our main focus is the two-channel model~\cite{Timmermans1999fri}, which is a standard tool to describe Fano-Feshbach resonances in ultracold atomic gases~\cite{Chin2010}. Here, the interaction proceeds via the coupling of the entrance channel to a closed channel (in realistic interactions, these channels are typically characterized by different spin configurations). As a result, the closed channel can only be occupied by one boson
at a time, which leads to an effective three-body repulsion. We show that we can transform the two-channel model in such a way that it allows an analytic solution of the three-body problem (i.e., $N=2$). In particular, we solve the atom-dimer scattering problem analytically and we predict the existence of a single three-body bound state (trimer) when the scattering length $a$ exceeds a critical value $a^*$ --- see Fig.~\ref{fig:mass}(b,d).
Our effective model furthermore allows the straightforward numerical calculation of the four-body bound state (tetramer) energy. Again, we predict the existence of a single bound state that, remarkably, only exists when $a>a^*$. We explain this fact
by noting that the effective model of $N-1$ bosons and one infinitely heavy dimer close to $a^*$ can be mapped exactly onto the problem of $N$ bosons and one infinitely heavy impurity close to unitarity~\cite{shortpaper}. For this reason, $a^*$ is also a multi-body resonance. 
The universal nature of our results is confirmed by considering the three- and four-body bound state spectrum within a second model, where an effective three-body repulsion is introduced by imposing a high-energy cutoff on scattering processes involving the exchange of two bosons~\cite{Bedaque1999}.

The paper is organized as follows. In Sec.~\ref{sec:models} we
derive the effective model that we use for our analysis of the
$(N+1)$-body problem. Starting from a two-channel
Hamiltonian for a single impurity atom, 
we map it to an Anderson-like model and then derive an effective Hamiltonian where the impurity degrees of freedom are essentially integrated out of the problem.
This approach forms the basis for our analytic solution of the
three-body problem presented in Sec.~\ref{sec:3body}. In
Sec.~\ref{sec:4body} we present our results for the
four-body problem, and in Sec.~\ref{sec:Nbody} we discuss how our
results generalize to an arbitrary number of bosons. In Sec.~\ref{sec:conc} we
conclude.

\section{Effective models}
\label{sec:models}

In this section, we derive an effective Hamiltonian that allows us to solve the three-body problem analytically. 
To this end, we start by introducing the two-channel model for a single impurity, which includes
a closed-channel molecule arising from the coupling of a boson and the impurity, and is known to faithfully reproduce the physics of Feshbach resonances~\cite{Timmermans1999fri}.
We then map this to a 
model which has a structure similar to the Anderson impurity model~\cite{anderson1961}, except that we are considering scalar bosons instead of spin-$1/2$ fermions.
Finally, we reach the desired effective model 
by diagonalizing the bilinear part in the bosonic Anderson model. For completeness, we also introduce the $\Lambda$-model, which we will use to test the universality of our results.

The final effective model only has the bosonic degrees of freedom, and the presence of the impurity is not manifest.
However, we will consistently refer to the scenario of one impurity and $N$ bosons as an $(N+1)$-body system such that we can use the same terminology within all models.

\subsection{Two-channel model for a single impurity}

We consider the two-channel Hamiltonian~\cite{Timmermans1999fri}
\begin{align}
    H_\mathrm{2ch} &=
    \sum_{\k} \epsilon_\k b_\k^\dag b_\k
    + \nu_0 d^\dag d 
    + g \sum_{\k} \left(
        d^\dag c\, b_\k
        + b_\k^\dag c^\dag d
    \right),
    \label{eq:2ch}
\end{align}
where $\epsilon_\k=k^2/2m$, $m$ is the mass of a boson, and $b_\k^\dag$, $c^\dag$, and $d^\dag$ are the creation operators of bosons, the impurity, and the closed-channel molecule, respectively. Throughout this paper, we work in units where the system volume and $\hbar$ are both set to~1.
The impurity, as well as the closed-channel molecule, is assumed to be localized at the origin, which thus allows us to drop the momentum dependence. 
The bare detuning $\nu_0$ and the coupling strength $g$ of a Feshbach resonance control the interaction, and are related to the scattering length $a$ and the effective range $r_0$ by
\begin{align}
    a\inv = \frac{2\pi}{mg^2} \left(
        \frac{mg^2}{\pi^2}k_0 - \nu_0
    \right), \quad
    r_0 = - \frac{2\pi}{m^2 g^2},
    \label{eq:renorm}
\end{align}
where $k_0$ is an ultraviolet (UV) cutoff which will be taken to infinity while keeping $a\inv$ and $r_0$ fixed.
Note that Eq.~\eqref{eq:renorm} implies that the effective range is always negative within this model.

Equation~\eqref{eq:renorm} fixes all physical two-body properties in the limit of $k_0\to \infty$.
The scattering of a boson by the impurity at total energy $E$ is described by the $T$ matrix
\begin{align}
    T(E+i0)=
    \frac{2\pi}{m}
    \frac{1}{a\inv - m r_0 E - \sqrt{-2mE-i0}},
    \label{eq:tmat}
\end{align}
which is independent of center-of-mass momentum in this model.
Here, $\pm 0$ represents a positive or negative infinitesimal number.
The $T$ matrix also gives the energy of a physical dimer as its pole.
It is known that for $a>0$, the two-channel model~\eqref{eq:2ch} has one diatomic bound state, or dimer, whose binding energy $\eb$ is
\begin{align}
    E_\mathrm{B} &\equiv \frac{\kappa_\mathrm{B}^2}{2m}
    = \frac{1}{2m r_0^2} \left(1-\sqrt{1-2\frac{r_0}a}
    \right)^2,
\end{align}
which we define to be positive. For later convenience, we have introduced the associated momentum $\kappa_\mathrm{B}$.
If, on the other hand, the scattering length is negative there is no two-body bound state.

\subsection{Bosonic Anderson model}

The restricted Hilbert space of a single impurity allows us to consider instead the model
\begin{align}
    H_\mathrm{A} &= H_{\mathrm{A}0} + H_{\mathrm{A}1},
    \label{eq:anderson}
    \end{align}
    with
    \begin{align}
    H_{\mathrm{A}0} &= \sum_{\k} \epsilon_\k b_\k^\dag b_\k
    + \nu_0 d^\dag d 
    + g \sum_{\k} \left(
        d^\dag b_\k
        + b_\k^\dag d
    \right), \label{eq:single}\\
    H_{\mathrm{A}1} &= \frac{U}{2} d^\dag d^\dag dd,
\end{align}
where we assume that the impurity operator $d^\dag$ is bosonic and
we take $U\to +\infty$ at the end of the calculation. Due to its formal similarity with the Anderson impurity model~\cite{anderson1961} we term the model~\eqref{eq:anderson} a ``bosonic Anderson model''.
Although $H_{\mathrm{A}0}$ looks similar to $H_\mathrm{2ch}$, it 
is effectively a single-particle Hamiltonian since the impurity degrees of freedom have been reduced to the operator $d^\dag$.
In the problem of two distinguishable particles, one can show that $H_\mathrm{2ch}$ and $H_{\mathrm{A}0}$ give the same results~\cite{Gurarie2007}.
The addition of the infinite on-site repulsion $H_\mathrm{A1}$ allows us to extend the equivalence to systems of two or more bosons interacting with a single infinitely heavy impurity.

The equivalence of $H_\mathrm{2ch}$ and $H_\mathrm{A}$ is most directly seen by comparing their corresponding \sch equations.
For simplicity, we present our arguments for the simplest non-trivial case of the three-body problem, involving the impurity and two bosons. However, the generalization to an arbitrary number of bosons is straightforward.
A general three-body state within the two-channel model takes the form
\begin{align}
    \ket{\Psi_\mathrm{2ch}}
    &= \!\! \left[ \frac{1}{2} \sum_{\k_1, \k_2} \psi_c(\k_1, \k_2)
            b_{\k_1}^\dag b_{\k_2}^\dag c^\dag \!
    + \sum_\k \psi_d(\k) b_\k^\dag d^\dag \right] \!\! \ket{0},
\end{align}
where the subscripts $c$ and $d$ of the wave functions indicate the presence of the particles denoted by these operators.
The wave function $\psi_c(\k_1, \k_2) $ is symmetric under the exchange of $\k_1$ and $\k_2$, reflecting the bosonic exchange symmetry.
Then one can write down the time-independent \sch equation $(E-H_\mathrm{2ch})\ket{\Psi_\mathrm{2ch}} = 0$ as follows:
\begin{subequations} \label{eq:sch-2ch-3body}
\begin{align}
    (E - \ekone - \ektwo) \psi_c(\k_1, \k_2) 
    &= g\left[ \psi_d(\k_1) + \psi_d(\k_2) \right], \\
    (E - \nu_0 - \ek) \psi_d(\k)
    &= g \sum_\q \psi_c(\k, \q).
\end{align}
\end{subequations}
On the other hand, a general three-body state for the bosonic Anderson model is
\begin{align}
    \ket{\Phi_\mathrm{A}}
    = & \left[
        \frac{1}{2} \sum_{\k_1, \k_2}
            \phi(\k_1, \k_2) b_{\k_1}^\dag b_{\k_2}^\dag 
    \right. \nn \\
    &\left. 
    \quad
        + \sum_{k} \phi_d(\k) b_\k^\dag d^\dag
        + \frac{1}{2} \phi_{dd} d^\dag d^\dag
    \right] \ket{0},
\end{align}
whose \sch equation reads
\begin{subequations} \label{eq:sch-A-3body}
\begin{align}
    (E - \ekone - \ektwo) \phi(\k_1, \k_2)
    &= g\left[ 
        \phi_d (\k_1) + \phi_d (\k_2)
    \right], \\
    (E - \nu_0 - \ek) \phi_d(\k)
    &= g\sum_\q \phi(\k, \q) + g \phi_{dd}, \\
    (E - 2\nu_0 - U) \phi_{dd}
    &= 2 g \sum_\q \phi_d(\q).
    \label{eq:phi_dd}
\end{align}
\end{subequations}
By comparing the sets of equations~\eqref{eq:sch-2ch-3body} and \eqref{eq:sch-A-3body}, one immediately sees that they are equivalent under the correspondence $\psi_c(\k_1,\k_2)=\phi(\k_1,\k_2)$ and $\psi_d(\k)=\phi_d(\k)$, provided $\phi_{dd}= 0$.
The latter condition is satisfied by taking $U\to \infty$ in Eq.~\eqref{eq:phi_dd}~\footnote{Here, we take $U\to \infty$ before we remove the UV cutoff in the integral.
Instead, one can take the infinite-cutoff limit first and the infinite-repulsion limit next. 
In that case, we can use Eq.~\eqref{eq:renorm} to subtract the UV divergence of the integral in Eq.~\eqref{eq:phi_dd}.
The limit $U\to\infty$ then leads to $\phi_{dd}=0$.}:
\begin{align}
    \phi_{dd}
    = \frac{2g}{E - 2\nu_0 - U} \sum_\q \phi_d(\q)
    \to 0.
\end{align}
Thus, we conclude that the two models give exactly the same results after taking $U\to \infty$.

\begin{comment}
We note that the models 
$H_\mathrm{2ch}$ and $H_\mathrm{A}$ are not connected by a unitary transformation.
This can be seen from the fact that the two-channel model conserves impurity number $N_\mathrm{i}=c^\dag c + d^\dag d$ while the bosonic Anderson model does not.
The conservation law, together with the non-negativity of $c^\dag c$, leads to the inequality $d^\dag d\leq 1$ in the two-channel model when $N_\mathrm{i}=1$.
In the bosonic Anderson model, on the other hand, the infinite on-site repulsion prohibits multiple occupancy of the localized closed-channel state, implementing the same inequality for the dimer state.
This implies that the bosonic Anderson model cannot be used in the presence of two impurities.
The correspondence of the two models, therefore, relies on the assumption that there is only one impurity.
\end{comment}  

The Hamiltonian~\eqref{eq:anderson} is also equivalent to the one-dimensional Anderson model with a $p$-wave coupling which, however, is not directly used in this paper.
See Appendix~\ref{ap:p-wave} for details.

\subsection{Derivation of the effective Hamiltonian}

The bosonic Anderson model~\eqref{eq:anderson} is nontrivial because of the channel mixing in $H_\mathrm{A0}$ and the on-site interaction in $H_\mathrm{A1}$.
Of these, we can eliminate the mixing by diagonalizing the bilinear Hamiltonian $H_\mathrm{A0}$.
This amounts to solving the two-body problem of the impurity and a boson; this solution is already known~\cite{braaten2008scattering}, but for completeness we outline it in the following.

The creation operator of a one-boson state within the model \eqref{eq:anderson} can be expanded in terms of $b_\p$ and $d$:
\begin{align}
    B_\k^\dag
    = \sum_\p
        \zeta_{\k\p} b_\p^\dag
    + \eta_{\k} d^\dag .
    \label{eq:Bop}
\end{align}
Here it is important to distinguish scattering states of the boson from the bound state that only exists when $a>0$. To this end, we let $\k \in \mathbb{R}^3\cup\{i\kappa_\mathrm{B}\}$ if $a>0$ and $\k\in\mathbb{R}^3$ if $a<0$, reflecting the presence and the absence of the physical two-body bound state. 
In both cases, the sum in Eq.~\eqref{eq:Bop} is over $\p\in \mathbb{R}^3$.
The conversion coefficients are the wave function of the two-body state,
\begin{align}
    \zeta_{\k\p}
    &= \begin{cases}
        \delta_{\k,\p} 
        + \frac{2m}{k^2 - p^2 + i0} T(\epsilon_\k+i0),
        & \quad \k \in \mathbb{R}^3, \\
        - \frac{2mg}{\kappa_\mathrm{B}^2+p^2}
        \eta_{i\kappa_\mathrm{B}},
        & \quad \k=i\kappa_\mathrm{B},
    \end{cases}
    \label{eq:zeta}
\end{align}
and
\begin{align}
    \eta_\k
    = \begin{cases}
        g\inv T(\epsilon_\k+i0),
            & \quad \k\in \mathbb{R}^3, \\
        \sqrt{\frac{\kappa_\mathrm{B}|r_0|}{\kappa_\mathrm{B}|r_0|+1}},
            & \quad \k = i\kappa_\mathrm{B}.
    \end{cases}
    \label{eq:eta}
\end{align}
We again stress that $\k=i\kappa_{\rm B}$ is only possible when $a>0$. $T(E)$ is the $T$ matrix given in Eq.~\eqref{eq:tmat}, and we have chosen the basis set consistent with the boundary condition of an incoming plane wave and an outgoing scattered wave, i.e., we use a positive infinitesimal imaginary shift of the energy.
The unitarity of the transformation \eqref{eq:Bop} is straightforward to verify
by direct calculation.

By rewriting $H_\mathrm{A}$ in terms of the new operators $B_\k$, we obtain the effective model upon which our analysis below is based:
\begin{align}
    H 
    ={\sum_{\k}}'\epsilon_\k B_\k^\dagger B_\k+\frac{U}{2} {\sum_{\substack{\k,\p\\ \u,\v}}}'
        \chi_{\k \p}^* \chi_{\u \v}
        B_{\k}^\dag B_{\p}^\dag B_{\u} B_{\v}.\label{eq:effmodel}
\end{align}
Here, ${\sum}'_\k h(\k)$ is a short-hand notation for
\begin{align}
    {\sum}'_\k h(\k) & = \sum_{\k\in\mathbb{R}^3} h(\k) + \begin{cases}  h(i\kb), &\quad a>0,\\
0, &\quad a\leq0.
\end{cases}
\end{align}
The coupling function $\chi_{\k\p}$ is further separable as it is given by
\begin{align}
    \chi_{\k\p} = \eta_\k \eta_\p.
    \label{eq:chi}
\end{align}
As a consequence, our Hamiltonian in Eq.~\eqref{eq:effmodel} is effectively one-dimensional, since the function $\eta$ defined in Eq.~\eqref{eq:eta} is independent of the direction of momentum. This observation greatly simplifies the investigation of the $(N+1)$-body problem below. However, for simplicity we will consistently use a notation appropriate for three spatial dimensions.

Our final Hamiltonian~\eqref{eq:effmodel} resembles that of interacting spinless bosons, since it consists of a diagonal quadratic term and a quartic interaction.
However, 
we emphasize that the two-boson interaction in Eq.~\eqref{eq:effmodel} is essentially of a three-body nature 
since it is induced by the presence of the impurity.
Specifically, when a boson occupies the closed-channel dimer state at the impurity's position, an additional boson is blocked from entering the closed channel and thus experiences an effective three-body repulsion.
Hence, the derivation of our effective Hamiltonian Eq.~\eqref{eq:effmodel} 
highlights
the three-body nature of the two-channel model, which is implicit in the original Hamiltonian~\eqref{eq:2ch}.

\subsection{$\Lambda$-model}
\label{sec:lambda}

In this work, we focus our attention on the physically realistic
two-channel model, i.e., on the effective Hamiltonian,
Eq.~\eqref{eq:effmodel}. However, in order to test the
model-independence of our results, we also show results calculated
using an alternative model,
which we term the $\Lambda$-model~\cite{Yoshida2018}. Here we take the
coupling $g\to\infty$ while keeping $a$ fixed in such a manner that the effective
range $r_0=0$. Thus, the two-body physics is effectively described within a
single-channel model where, e.g., the dimer binding energy is simply
$\eb=1/2ma^2$. However, to regularize the few-body problem beyond
two-body, we introduce a cutoff $\Lambda$ on all momentum
sums in the few-body equations resulting from an exchange of
bosons. In the three-body problem, this procedure is known to be
equivalent to introducing an effective three-body
repulsion~\cite{Bedaque1999}. We also note that in Ref.~\cite{Yoshida2018} it was shown
that the two-channel model and the
$\Lambda$-model have --- to a very high degree of accuracy --- the same universal
few-body physics for an impurity of mass $M=m$, once $r_0$ and
$\Lambda$ are both related to the three-body parameter $a_-$. For more
details on this model, see Appendix~\ref{ap:lambda}.

\section{Exact solution of the three-body problem}
\label{sec:3body}

We now turn to the three-body problem consisting of two non-interacting bosons and an infinitely heavy impurity. 
This scenario may be regarded as an extreme limit of the quantum three-body problem, which has a 
long and celebrated history starting with 
Efimov,
who first predicted that three identical bosons 
can support infinitely many bound states near a two-body resonance~\cite{Efimov1970}. 
Efimov's original scenario 
has since been generalized to other three-body systems to study the effect of particle statistics~\cite{Efimov1973,Braaten2006,helfrich2010,Yoshida2018} 
and finite-range interactions~\cite{Petrov2004tbp,Levinsen2011,Wang2012a,naidon2014a}. In the case of two heavy fermonic atoms and one light particle, it was shown by Efimov himself that the Efimov effect only occurs when the mass ratio is above a critical value, $m/M\geq 13.6\ldots$~\cite{Efimov1973}. For mass ratios lower than $13.6$, Kartavtsev and Malykh pointed out that the system can support another type of trimer state for positive scattering length~\cite{kartavtsev2007low}: Unlike the Efimov trimers, these trimers are universal in the sense that their energies do not depend on the short-range details of the interaction (i.e. a three-body parameter)~\cite{jag2014observation,shi2014universal,shi2015efimov,kartavtsev2016universal}. However, later studies of finite-range systems show that 
the universal behavior is only limited to very large scattering lengths~\cite{endo2012,safavi2013nonuniversal}. This suggests that even for systems with a mass ratio lower than the critical value, there are still remnants of the Efimov effect since the three-body parameter can have a visible impact. 
Indeed, in Ref.~\cite{Gao2015}, the authors found that the three-body parameter could modify the third virial coefficient $b^{(2,1)}$ for mass ratios below $13.6$.

For our three-body problem of one fixed impurity and two bosons,
although there is no Efimov physics here, the system is right at the edge of the Efimov region, since the Efimov effect occurs for any finite impurity mass (Fig.~\ref{fig:mass}). 
Therefore, it is crucial to include a high-energy three-body parameter (effective range $r_0$ or momentum cutoff $\Lambda$) in our models to capture the short-range physics.

As we now discuss, the separability of the interaction within our effective Hamiltonian~\eqref{eq:effmodel}
enables the exact treatment of the three-body problem.
Here, focusing on the case with a positive scattering length, where the two-body problem has scattering states as well as a two-body bound dimer state, we investigate the trimer state and the scattering between the dimer and a boson.
We consider a wave function in the basis spanned by single-boson states $B_\k^\dag$~\eqref{eq:Bop} as follows:
\begin{align}
    \ket{\Phi} 
    = \frac{1}{2} {\sum_{\p,\q}}' 
        \varphi_{\p\q} B_\p^\dag B_\q^\dag \ket{0}.
        \label{eq:3body}
\end{align}
Here, the wave function $\varphi_{\p\q}$ is symmetric under the exchange of $\p$ and $\q$.
It satisfies the \sch equation $(E-H) \ket{\Phi}=0$, which reads
\begin{align}
    (E - \epsilon_\p - \epsilon_\q) \varphi_{\p\q}
    = U {\sum_{\u,\v}}' \chi_{\p\q}^\ast \chi_{\u\v} \varphi_{\u\v}.
    \label{eq:sch3body}
\end{align}
Note that the double summation on the right-hand side of this equation 
contains a double integral on $\u,\v\in \mathbb{R}^3$, single integrals with $\u=i\kappa_\mathrm{B}$ or $\v=i\kappa_\mathrm{B}$, and a term with $\u=\v=i\kappa_\mathrm{B}$.
In the following sub-sections, we analyze this equation in the case of $E<-E_\mathrm{B}$ (three-body bound state) and for $-E_\mathrm{B} \leq E < 0$ (atom-dimer scattering).

\subsection{Bound state}

Assuming that the right-hand side of Eq.~\eqref{eq:sch3body} is non-zero in the limit of $U\to \infty$, we can invert the equation to obtain
\begin{align}
    \varphi_{\p\q}
    = \frac{\chi_{\p\q}^\ast f}{E-\epsilon_\p-\epsilon_\q},
    \label{eq:trimer_wf}
\end{align}
where $f\equiv U{\sum}'_{\p,\q} \chi_{\p\q} \varphi_{\p\q}$.
By substituting Eq.~\eqref{eq:trimer_wf} back into the definition of $f$, we find
\begin{align}
    0
    &= \frac{1}{U} 
        - {\sum_{\p,\q}}' \frac{|\chi_{\p\q}|^2}{E - \epsilon_\p - \epsilon_\q} 
    \equiv \frac{1}{U} - Z(E),
    \label{eq:trimereng}
\end{align}
where we have defined the three-body function $Z(E)$ and we have used the assumption that $f\neq 0$.
From Eq.~\eqref{eq:trimereng}, one can see that after taking $U\to\infty$, the energy $E_3$ of a trimer, if any, is found as a zero of $Z(E)$ on the real axis.

Before evaluating $Z(E)$, we can draw some conclusions from its expression.
First, the trimer energy satisfies $E>- 2 E_\mathrm{B}$, since each term of the summation in Eq.~\eqref{eq:trimereng} is negative when $E<-2\eb$ 
and hence it is not possible to satisfy $Z(E)=0$.
Second, there is at most one bound trimer state. This is a consequence of each term in $\partial Z/\partial E$ being negative when $-2E_\mathrm{B} < E < -E_\mathrm{B}$, implying that $Z(E)$ is monotonic throughout the range of energies where a trimer is possible.
Third, $Z(E)$ is singular at both ends of this energy interval:
At $E=-2E_\mathrm{B}$, it has a simple pole that stems from the two-dimer state, while 
 at $E=-E_\mathrm{B}$, there is the termination of a branch cut that corresponds to
the atom-dimer continuum states.

By performing the contour integrals in Eq.~\eqref{eq:trimereng}, we find the analytic expression for $Z(E)$: 
\begin{widetext} 
\begin{align} 
    Z\left( -\frac{\kappa^2}{2m} \right)
    = \frac{2m}{\kappa^2(\cos\theta_+ - \cos\theta_-)^2} \left\{
    \frac{3\cos\theta_+(\cos\theta_+ + \cos\theta_-)}
        {(\sin\theta_+ + \cos\theta_+)(\sin\theta_+ - \cos\theta_-)}
    - \frac{\cos\theta_- (\cos\theta_+ + \cos\theta_-)}
        {(\sin\theta_- + \cos\theta_+)(\sin\theta_- - \cos\theta_-)} \right. \nn\\
    \left. + \frac{4\cos^2\theta_+}{\cos 2\theta_+}
    + \frac{\cos^2\theta_+ - \cos^2\theta_-}{\cos2\theta_+ + \cos2\theta_-}
    \left[
        \left( \frac{4\theta_+}{\pi} - 1 \right) \tan 2\theta_+
        - \left( \frac{4\theta_-}{\pi} - 3 \right) \tan 2\theta_-
    \right]
    \right\}.
    \label{eq:Z}
\end{align} 
\end{widetext}
Here, we define $\theta_\pm \equiv \arccos \frac{\kappa_\pm}{\kappa}$, where the poles $\p=i\kappa_\pm$ of $T(\epsilon_\p)$ are $\kappa_\pm\equiv(\pm\sqrt{1+2|r_0|/a}-1)/|r_0|$.
We take the branch of $\arccos x$, where $0<\arccos x<\pi$ for $-1\leq x\leq 1$ and $\arccos x=\pi - i\,\mathrm{arccosh}\, |x|$ for $x < -1$.
Note that the positive imaginary pole $\kappa_+=\kappa_\mathrm{B}$ corresponds to the two-body bound state, whereas the negative imaginary pole $\p=i\kappa_-$ does not correspond to a physical state.

\begin{figure}
    \centering
    \includegraphics[width=0.9\columnwidth]{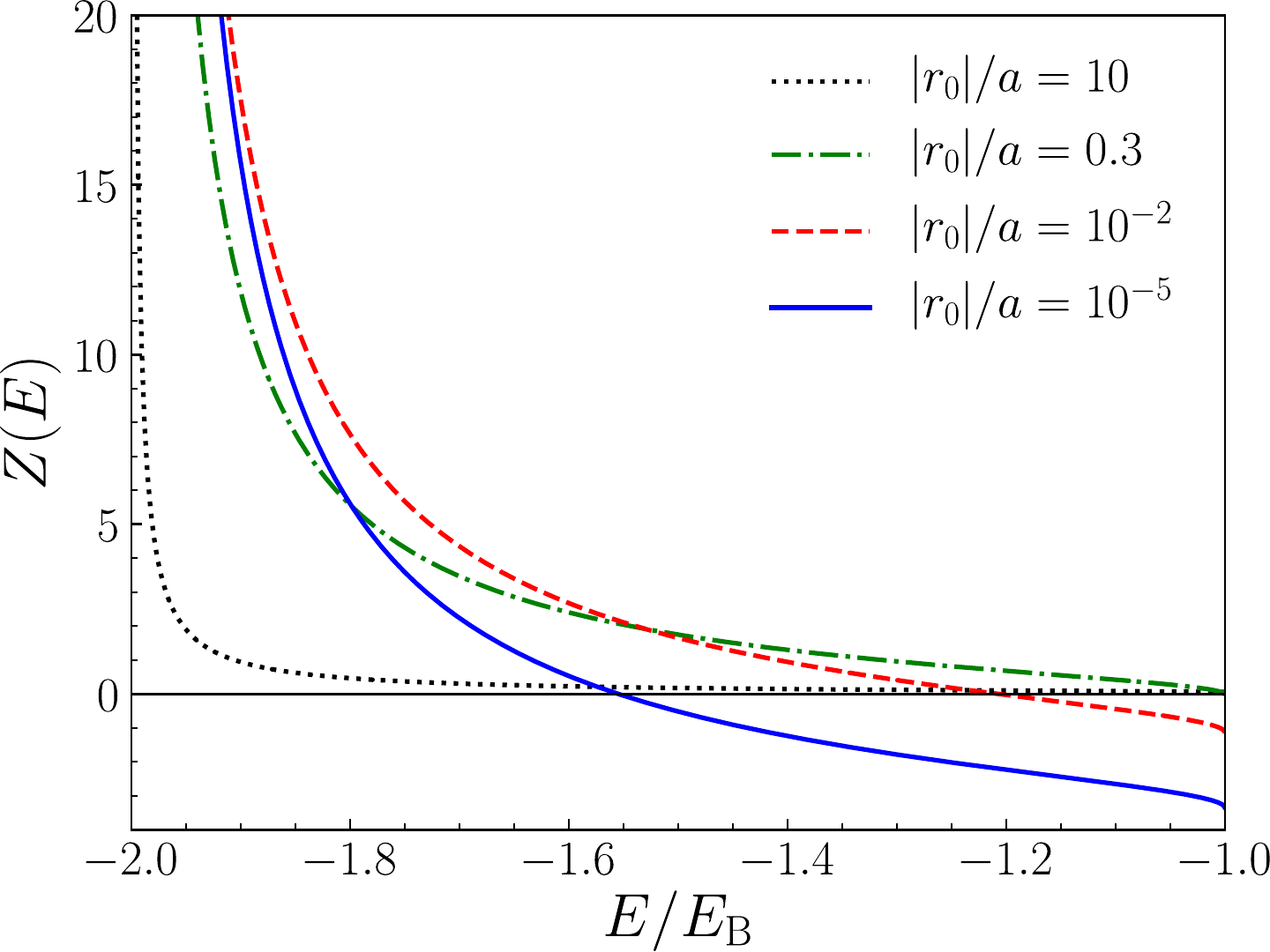}
    \caption{The three-body function $Z(E)$ within the interval $-2E_\mathrm{B}<E<-E_\mathrm{B}$ for several values of $|r_0|/a$.
    For $a/|r_0| < a^*/|r_0| \simeq 1/0.31821\dots$, $Z(E)$ is positive within the plotted energy range, implying that no trimer exists in this range when $U\to\infty$.}
    \label{fig:zfunc}
\end{figure}
In Fig.~\ref{fig:zfunc} we plot the function $Z(E)$ in the range of energies $-2E_\mathrm{B}<E<-E_\mathrm{B}$,
which illustrates how $Z(E)$ is a monotonic function with a simple pole at $E=-2E_\mathrm{B}$, as described above. 
Since the trimer energy $E_3$ corresponds to the root $Z(E_3)=0$, this can immediately be read off for a given $|r_0|/a$.
Close to resonance where $|r_0|/a\to +0$ we see how the trimer energy only very slowly approaches $-2E_\mathrm{B}$.
On the other hand, as $|r_0|/a$ is made larger, $Z(-E_\mathrm{B})$ increases until finally it crosses zero at a critical scattering length $a^*\simeq |r_0|/0.31821$ implying that, for an even larger $|r_0|/a$,
the trimer state ceases to exist. The parameter $a^*$ is equivalent to the critical scattering length at which the ground state in the Efimov trimer spectrum crosses into the atom-dimer continuum~\cite{Efimov1970}.

According to Eq.~\eqref{eq:trimereng}, $Z(E)$
actually provides the trimer energy even if we were to extend the model to arbitrary $U$; in that case, the energy would correspond to the crossing of $Z(E)$ in Fig.~\ref{fig:zfunc} with a horizontal line at $1/U$. In particular, from Fig.~\ref{fig:zfunc} one can see that, for any $|r_0|/a$, the trimer energy converges to $-2E_\mathrm{B}$ as $U\to +0$. This is the result anticipated for two non-interacting bosons in a static potential that has a bound state with energy $-\eb$, and thus corresponds to the case of no induced boson-boson repulsion.

\begin{figure}
    \centering
    \includegraphics[width=0.9\columnwidth]{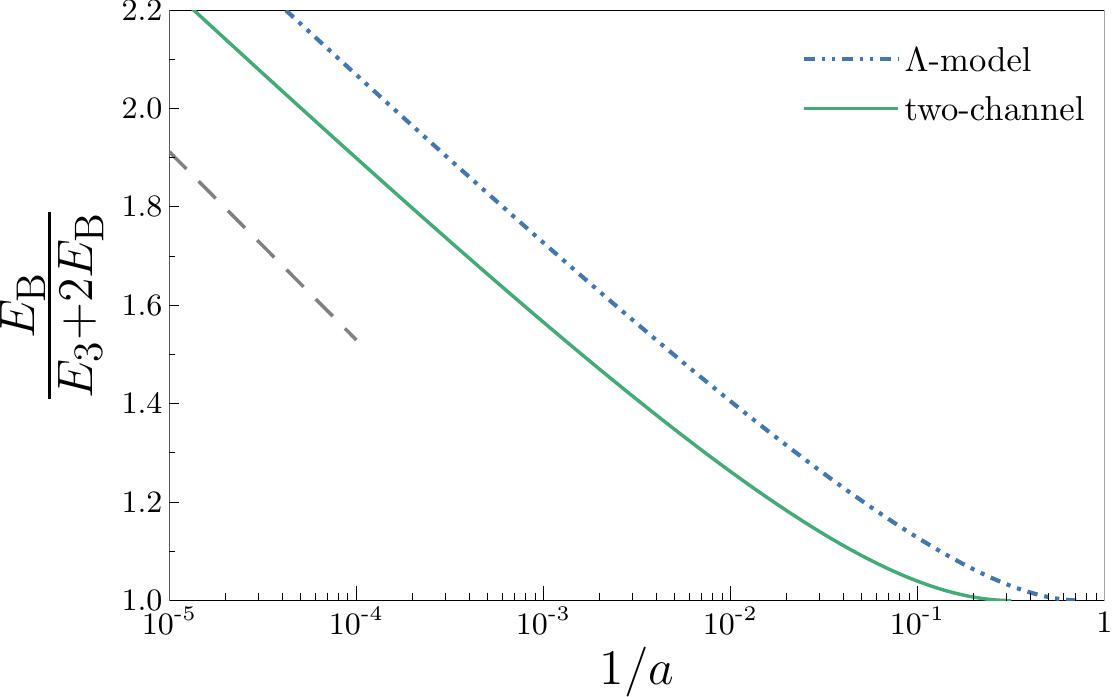}
    \caption{Inverse correction $\frac{E_\text{B}}{E_3+2E_\text{B}}$ as a function of $1/a$ for the two-channel model (green solid) and $\Lambda$-model (blue dash-dotted). The scattering length $a$ is in units of either $|r_0|$ or $1/\Lambda$ for the different models. The dashed line is a guide to the eye with slope $1/2\pi$.}
    \label{fig}
\end{figure}

We can analyze our results further in the limit
$|r_0|/a \to 0$, 
where Eq.~\eqref{eq:Z} reduces to
the asymptotic expression
\begin{align}
    Z \left( -\frac{\kappa^2}{2m}\right) 
    \simeq \frac{8m}{\kappa_-^2} \left(
        \frac{\kappa_+^2}{2\kappa_+^2 - \kappa^2}
        - \frac{1}{2\pi} \log \frac{|\kappa_-|}{\kappa}
    \right),
\end{align}
or equivalently,
\begin{align}
    Z \left( E\right) 
    \simeq 2mr_0^2 \left[
        \frac{E_\mathrm{B}}{2E_\mathrm{B} + E}
        + \frac{1}{4\pi} \log \left( mr_0^2E \right)
    \right].
    \label{eq:Z-asymp}
\end{align}
Since $mr_0^2 E\to 0$ near the resonance, this equation implies that $E\to -2E_\mathrm{B}\simeq -1/ma^2$ to compensate for the logarithmic divergence.
Therefore, we can find an approximate root of $Z(E)$:
\begin{align}
    E_3 \simeq -2 E_\mathrm{B} + \frac{2\pi E_\mathrm{B}}{\log (a/|r_0|)}.
    \label{eq:3Basymp}
\end{align}
The first term is consistent with the result for two non-interacting bosons in a static potential that has a bound state with energy $-E_\mathrm{B}$. However, we see that for the realistic interatomic interactions described by the two-channel model, there is a strong correction to this result that only decays logarithmically. 

We emphasize that the logarithmic correction found in Eq.~\eqref{eq:3Basymp} is essentially independent of $r_0$, if we regard the limit $|r_0|/a\to0$ as corresponding to $a\to +\infty$ while keeping $r_0$ fixed.
More precisely, we can replace $|r_0|$ in Eq.~\eqref{eq:3Basymp} with any length scale without invalidating the expression up to the first logarithmic correction, because a change in the relative length scale only causes a constant shift of the logarithmic function.
This indicates that the scattering length is the only relevant parameter near the unitarity limit, and suggests a universality of the logarithmic correction.

We demonstrate this point numerically by calculating the trimer energy within the $\Lambda$-model and comparing the results with those of the two-channel model, where the latter corresponds to finding the roots of $Z(E)$ in Eq.~\eqref{eq:Z}.
By plotting the quantity $E_\mathrm{B}/(E+2E_\mathrm{B})$ for both models, as shown in Fig.~\ref{fig},
we see that both curves are asymptotically linear in $\log a$ as $a\to\infty$, with a slope consistent with that predicted by Eq.~\eqref{eq:3Basymp}.
We thus conclude that the energy of the trimer is a universal function of the scattering length up to the first logarithmic correction arising from the effective three-body interaction.

\subsection{Atom-dimer scattering}

We now consider the scattering of a boson and the dimer formed from a boson and the infinitely heavy impurity.
We focus on the energy range of $-E_\mathrm{B} < E< 0$, where only elastic processes occur.
In this case, the solution of Eq.~\eqref{eq:sch3body} is the sum of  terms representing the incoming wave with a fixed momentum $\q$ and the scattered wave:
\begin{align}
    \varphi_{\k\p}
    = 
    \delta_{\k,\q} \delta_{\p,i\kappa_\mathrm{B}}
    + \delta_{\p,\q} \delta_{\k,i\kappa_\mathrm{B}}
    + \frac{t_3(\k,\p;\q)}
        {E - \epsilon_\k - \epsilon_\p + i0},
    \label{eq:ad-wf-B-basis}
\end{align}
where we have introduced the amplitude $t_3(\k,\p;\q)$ of the scattered wave in the scattering-state basis:
\begin{align}
    t_3(\k,\p;\q)
    = U {\sum_{\u,\v}}' \chi_{\k\p}^\ast \chi_{\u\v} \varphi_{\u\v}.
\end{align}
Here, $\q$ is the incoming momentum and the total energy is
$E=-E_\mathrm{B}+\epsilon_\q$.
Note that the ``incoming wave'' in this notation does not describe a plane wave of a free boson and a dimer; in terms of the operators, such a state is written as $b^\dag_\q B^\dag_{i\kappa_\mathrm{B}} \ket{0}$.
In the current formalism, we adopt the basis spanned by $B_\q^\dag$ instead of $b_\q^\dag$, and therefore the incoming state is $B^\dag_\q B^\dag_{i\kappa_\mathrm{B}} \ket{0}$.
For the same reason, $t_3(\k,\p;\q)$ itself is not the $T$ matrix of the atom-dimer scattering, though they are related as we see in the next section.

We obtain a Lippmann-Schwinger-like equation for $t_3(\k,\p;\q)$ by plugging the wave function \eqref{eq:ad-wf-B-basis} into the definition of $t_3(\k,\p;\q)$:
\begin{align}
    t_3(\k,\p;\q)
    = 2 U \chi_{\k\p}^\ast \chi_{\q i\kappa_\mathrm{B}}
    + U {\sum_{\u,\v}}' 
        \frac{\chi_{\k\p}^\ast \chi_{\u\v}t_3(\u,\v;\q)}
        {E - \epsilon_\u - \epsilon_\v + i0}.
    \label{eq:LS}
\end{align}
Due to the separability of the interaction, the dependence on $\k,\p$ on the right-hand side is completely determined by $\chi_{\k\p}^\ast$.
We can thus substitute $\chi_{\k\p}^\ast c(\q)$ for $t_3(\k,\p;\q)$ and reduce the above integral equation into a linear equation for $c(\q)$. This procedure yields the following solution:
\begin{align}
    t_3(\k,\p;\q) 
    &= \frac{2\chi_{\k\p}^\ast \chi_{\q i\kappa_\mathrm{B}}}{U\inv - Z(E+i0)}
    \stackrel{U\to\infty}{\longrightarrow} -\frac{2\chi_{\k\p}^\ast \chi_{\q i\kappa_\mathrm{B}}}{Z(E+i0)}.
\end{align}
We thus obtain the full wave function of an atom-dimer scattering state.

\subsubsection{Scattering amplitude}

Information on the atom-dimer scattering, or the scattering amplitude, is encoded in the asymptotic wave consisting of a free boson and a physical dimer.
The wave function of relative atom-dimer motion is given in the free-boson basis as
\begin{align}
    \varphi_\mathrm{ad}(\k)
    \equiv \bra{0} b_\k B_{i\kappa_\mathrm{B}} \ket{\Phi}
    ={\sum_{\p}}'
        \varphi_{\p i\kappa_\mathrm{B}} \zeta_{\p\k},
    \label{eq:ad-wf}
\end{align}
where we have used the definition of the general three-body wave function $\ket{\Phi}$ in Eq.~\eqref{eq:3body}, as well as the definition of the $B$ operator, Eq.~\eqref{eq:Bop}.
This characterizes the direct process, where the first boson is free while the second one is bound to the impurity in both the initial and final states.
Although there is of course an exchange process due to the bosonic symmetry, $\varphi_\mathrm{ad}(\k)$ suffices for the purpose of deriving the scattering amplitude.
It contains a delta function corresponding to the incoming wave and a pole near the real axis corresponding to the outgoing scattered wave:
\begin{align}
    \varphi_\mathrm{ad}(\k)
    \simeq \delta_{\k,\q} 
    + \frac{T_\mathrm{ad}(\q)}{E+E_\mathrm{B}-\epsilon_\k+i0}.
    \label{eq:ad-asymp}
\end{align}
Here, $T_\mathrm{ad}(\q)$ is the (on-shell) atom-dimer $T$ matrix, where we ignore regular terms and poles at imaginary $\k$ because they do not contribute to the long-distance behavior of the wave function.

To derive the atom-dimer $T$ matrix, one needs to perform the integral in Eq.~\eqref{eq:ad-wf}.
By collecting the relevant terms in Eq.~\eqref{eq:ad-wf}, we find an expression for $T_\mathrm{ad}(\q)$ in terms of $T(E)$ and $t_3(\k,\p;\q)$:
\begin{align}
    T_\mathrm{ad}(\q)
    &= T(\epsilon_\q) 
    + t_3(\q, i\kappa_\mathrm{B};\q) \left[
        1 - \frac{imq}{\pi} T(\epsilon_\q)
    \right] \nn \\
    &= T(\epsilon_\q) 
    + e^{2i\delta_q} t_3 (\q,i\kappa_\mathrm{B};\q),
\label{eq:Tad}
\end{align}
where $\delta_q$ is the scattering phase shift of the atom-impurity scattering: $T(\epsilon_\q)=-\frac{2\pi}{mq} e^{i\delta_q} \sin\delta_q$. From Eq.~\eqref{eq:Tad} we see that the atom-dimer scattering amplitude is the sum of the bare atom-impurity scattering amplitude and a three-body term. The latter is a direct consequence of the impurity-induced repulsion between bosons, and it would disappear were we to instead take the limit $U\to0$ [see Eq.~\eqref{eq:LS}].

The scattering amplitude is the on-shell $T$ matrix multiplied by $-m/2\pi$.
Therefore, the atom-dimer scattering amplitude is
\begin{align}
    f_\mathrm{ad}(\q)
    = f_0 (\q) + e^{2i\delta_q} f_3(\q),
    \label{eq:f_ad}
\end{align}
where $f_0(\q)$ and $f_3(\q)$ are the scattering amplitudes corresponding to $T(\epsilon_\q)$ and $t_3(\q,i\kappa_\mathrm{B};\q)$, respectively.
By noting that 
\begin{align}
\mathrm{Im}[Z(\epsilon_\q-E_\mathrm{B}+i0)]=-\frac{mq}{\pi} |\chi_{\q i\kappa_\mathrm{B}}|^2
\label{eq:Im-Z}
\end{align}
for $q<\kappa_\mathrm{B}$, we obtain
\begin{align}
    f_3(\q)
    &= -
        \frac{\mathrm{Im}[Z(\epsilon_\q-E_\mathrm{B}+i0)]}
        {q Z(\epsilon_\q-E_\mathrm{B}+i0)}.
    \label{eq:f_3}
\end{align}
This expression for $f_3(\q)$ is consistent with the existence of a real phase shift $\delta_{3,q}$, which is related to $f_3(\q)$ by $f_3(\q)=q\inv e^{i\delta_{3,q}} \sin\delta_{3,q}$.
To make this point explicit, one can look at the inverse of the scattering amplitude: 
\begin{align}
    \frac{1}{q f_3(\q)}
    = \cot\delta_{3,q} - i
    = \frac{- \mathrm{Im}[Z(\epsilon_\q-E_\mathrm{B}+i0)]}
        {\mathrm{Re}[Z(\epsilon_\q-E_\mathrm{B}+i0)]}
    - i.
\end{align}
This indicates that $\cot\delta_{3,q}$ is real, which in turn implies that $\delta_{3,q}$ is real.

In addition, Eq.~\eqref{eq:f_ad} implies that
the atom-dimer phase shift $\delta_{\mathrm{ad},q}$ is given by
\begin{align}
    \delta_{\mathrm{ad},q}
    = \delta_q + \delta_{3,q}.
\end{align}
This can be shown by arranging the terms on the right-hand side of Eq.~\eqref{eq:f_ad} using the addition theorem of the trigonometric functions and comparing with $q^{-1}e^{i\delta_{\mathrm{ad},q}} \sin\delta_{\mathrm{ad},q}$.
The fact that $\delta_{\mathrm{ad},q}$ is real indicates that the atom-dimer scattering is elastic,
which is expected in the energy range under consideration. 

\begin{figure}
    \centering
    \includegraphics[width=0.9\columnwidth]{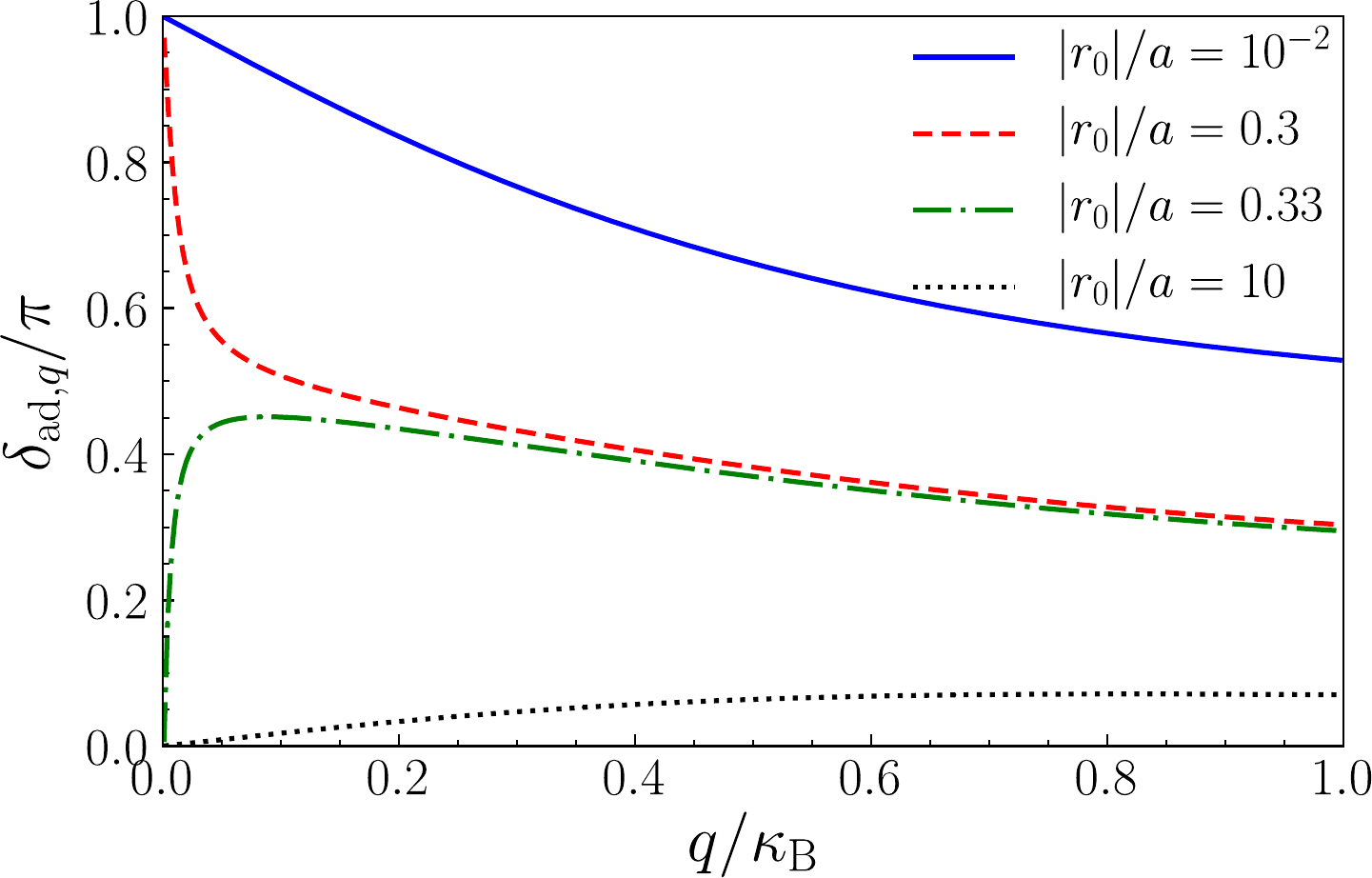}
    \caption{Phase shift for the atom-dimer scattering for several values of $|r_0|/a$.
    When $|r_0|/a = |r_0|/a^* \simeq 0.31821\dots$, the slope at $q=0$ diverges and changes sign.}
    \label{fig:phaseshift}
\end{figure}

In Fig.~\ref{fig:phaseshift}, we show the phase shift $\delta_{\mathrm{ad},q}$ for atom-dimer scattering.
Note that $\delta_{\mathrm{ad}, q}$ is defined modulo $\pi$.
One can extract the scattering length from this plot by using $\delta_{\mathrm{ad},q}\simeq -a_\mathrm{ad}q$ for small $q$. Therefore, the negative slopes for smaller $|r_0|/a$ imply that $a_\mathrm{ad}>0$, while the positive slope for $|r_0|/a=10$ indicates $a_\mathrm{ad}<0$.
In particular, the slope is large at $|r_0|/a=0.3$, which is close to the atom-dimer resonance at $a^*$
where $a_\mathrm{ad}\to \pm \infty$.
We now proceed to derive the analytical expressions for the low-energy scattering parameters, i.e., the atom-dimer scattering length and effective range, which explicitly demonstrate this behavior.

\subsubsection{Scattering length}

The most important parameter of low-energy atom-dimer scattering is the associated scattering length defined as $a_\mathrm{ad}=-f_\mathrm{ad}(0)$. 
From Eq.~\eqref{eq:f_ad} 
we have
\begin{align}
    a_\mathrm{ad}
    = a - f_3(0),
\end{align}
where we have used $e^{2i\delta_q}\to 1$ as $q\to0$.
By combining Eqs.~(\ref{eq:Im-Z},\ref{eq:f_3}), we find 
\begin{align}
    f_3(0)
    = \frac{m|\chi_{0 i\kappa_\mathrm{B}}|^2}{\pi Z(-E_\mathrm{B})}.
\end{align}
We can write $|\chi_{0 i\kappa_\mathrm{B}}|^2$ in terms of $a$ and $r_0$ via its definition \eqref{eq:eta} and \eqref{eq:chi} to obtain
\begin{align}
    a_\mathrm{ad} 
    &= a \left[
        1 - \frac{2m |r_0| a}{Z(-E_\mathrm{B})}
            \frac{\sqrt{1 + 2 |r_0| /a} - 1}{\sqrt{1 + 2 |r_0| /a}}
    \right].
\end{align}
This implies that $a_\mathrm{ad}$ diverges if either $Z(-E_\mathrm{B})\to0$ or if $a\to\pm\infty$.
This is an expected property of the atom-dimer scattering length.
The first condition is realized at $a=a^\ast$, where the trimer dissociates into the continuum of atom-dimer scattering states. On the other hand, the second condition corresponds to the unitarity limit, where both the trimer and dimer spectra merge into the free atom continuum.
Here, we find that
\begin{align}
    a_\mathrm{ad} \to a \left(
        1 + \frac{2\pi}{\log(a/|r_0|)}
    \right).
    \label{eq:aad-asymp}
\end{align}
This can be shown by taking the limit $|r_0|/a\to0$ and noting that  $Z(-E_\mathrm{B})/mr_0^2\simeq-\frac{1}{\pi}\log(a/|r_0|)$ from Eq.~\eqref{eq:Z-asymp}. The linear term again corresponds to the boson only interacting with the impurity, in a manner independent of the presence of another boson inside the dimer.

\begin{figure}
    \centering
    \includegraphics[width=0.9\columnwidth]{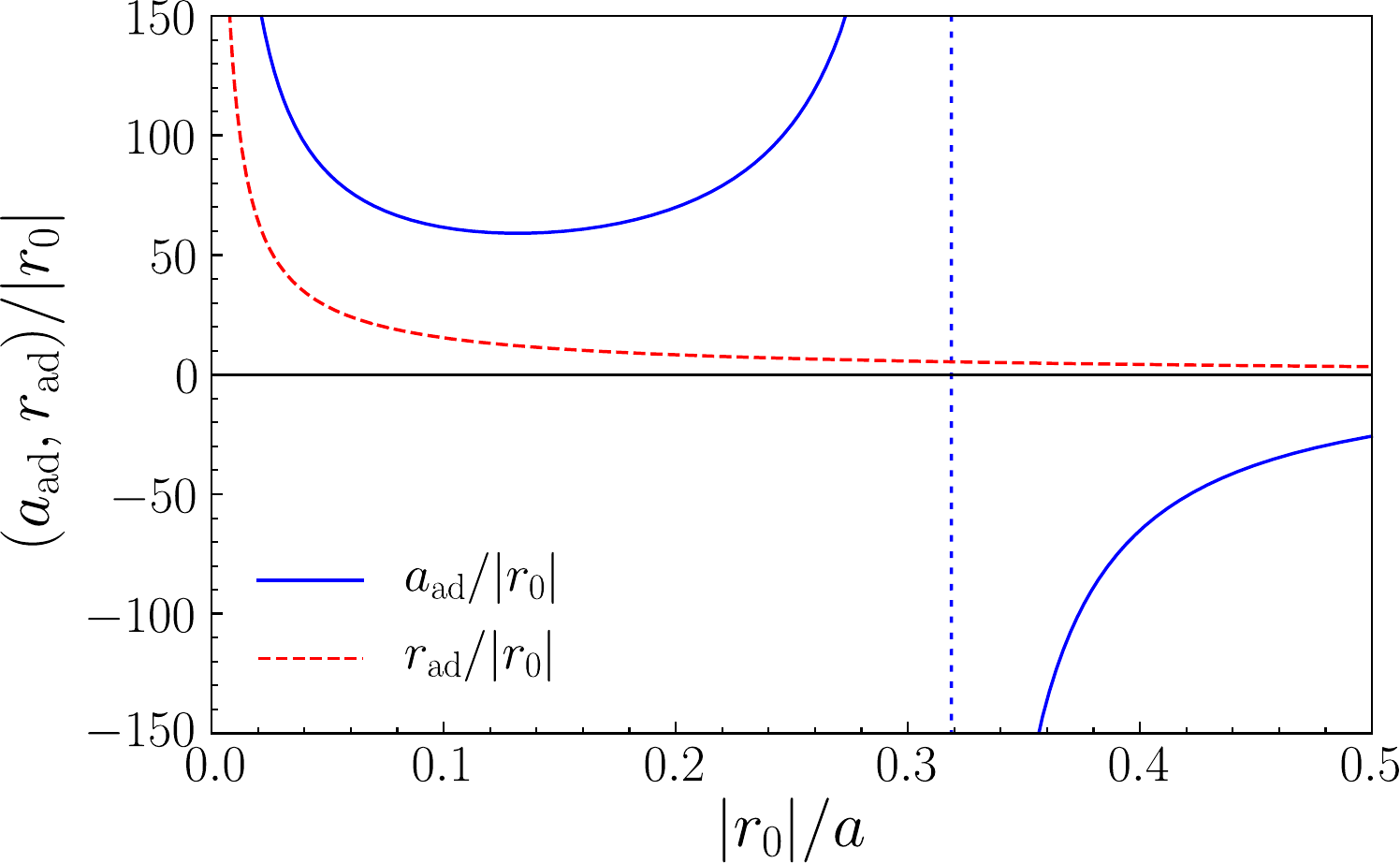}
    \caption{The scattering length (blue solid) and the effective range (red dashed) of the atom-dimer scattering.
    The scattering length diverges at $|r_0|/a = 0$ (the unitarity limit) and $0.31821$ (atom-dimer resonance), while $r_\mathrm{ad}$ is finite in the latter case.}
    \label{fig:aad-rad}
\end{figure}

We observe both of these features in Fig.~\ref{fig:aad-rad} which shows the atom-dimer scattering length as a function of $|r_0|/a$.
The figure also shows that $a_\mathrm{ad}$ is much larger than $|r_0|$ when $a_\mathrm{ad}>0$.
This is consistent with our observation that the trimer is generally very weakly bound relative to the dimer, except exponentially close to unitarity.

\subsubsection{Effective range}

The effective range $r_\mathrm{ad}$ of the atom-dimer scattering can be extracted from the low-energy expansion of the scattering amplitude:
\begin{align}
    f_\mathrm{ad}(\q)
    \simeq -a_\mathrm{ad} + ia_\mathrm{ad}^2 q
    + \left( 
        a_\mathrm{ad}^3 
        - \frac{a_\mathrm{ad}^2 r_\mathrm{ad}}{2}
    \right) q^2 .
    \label{eq:low-eng-amplitude}
\end{align}
To proceed, we relate $f_\mathrm{ad}(\q)$ to $f_0(\q)$ and $f_3(\q)$ using Eq.~\eqref{eq:f_ad}. Here we note that we already know the low-energy expansions of $f_0(\q)$ and $e^{2i\delta_q}$ in terms of $a$ and $r_0$.
To find $r_\mathrm{ad}$, therefore, the remaining ingredient is the expression of $f_3(\q)$ for $\eq/\eb\ll1$. From Eq.~\eqref{eq:f_3} this is seen to be equivalent to extracting the low-energy behavior of $Z(\epsilon_\q-E_\mathrm{B})$.

By inspecting its definition, one can see that
\begin{align}
    Z(\epsilon_\q-E_\mathrm{B})
    = z_0 + iz_1 q + z_2 q^2 + O(q^3),
\end{align}
where $z_0, z_1$, and $z_2$ are real coefficients.
This expansion, combined with those of $f_3(\q)$, $f_0(\q)$, and $e^{2i\delta_q}$, yields
\begin{align}
    r_\mathrm{ad}
    &= \frac{a^2r_0}{a_\mathrm{ad}^2} 
    - 2 \left(
        \frac{a}{a_\mathrm{ad}} - 1
    \right) \left(
        a - \frac{z_2}{a_\mathrm{ad} z_0}
    \right).
\end{align}
We already know the expressions for $z_0\equiv Z(-E_\mathrm{B})$ and $a_\mathrm{ad}$ in terms of $a$ and $r_0$.
We can also find $z_2$ by straightforward (though tedious) calculation:
\begin{align}
    &2m E_\mathrm{B}^2 z_2 
    = \frac{8(\alpha+1)(3\alpha^2-2)}{\pi\alpha^2(\alpha-1)(2\alpha^2-1)} \nn \\
    &- \frac{4(3+2\alpha-12\alpha^2-7\alpha^3+12\alpha^4+5\alpha^5+4\alpha^7)}
        {\alpha^3(\alpha-1)^2(2\alpha^2-1)^2} \\
    &- \frac{8\sqrt{\alpha^2-1}(2-7\alpha^2+4\alpha^4)\log(-\alpha+\sqrt{\alpha^2-1})}
        {\pi \alpha^3 (\alpha-1)^2(2\alpha^2-1)^2 }, \nn
\end{align}
where 
\begin{align}
    \alpha \equiv \frac{\kappa_-}{\kappa_+}
    = - \frac{\sqrt{1+2|r_0|/a} + 1}{\sqrt{1+2|r_0|/a} - 1}.
\end{align}

We show the resulting atom-dimer effective range in Fig.~\ref{fig:aad-rad}.
From the analytic expression, we find that it diverges with the scattering length as:
\begin{align}
    r_\mathrm{ad} \to \frac{4\pi a}{\log(a/|r_0|)}, \quad a\to\infty.
\end{align}
On the other hand, $r_\mathrm{ad}$ remains finite at the atom-dimer resonance.
Therefore, since 
$r_\mathrm{ad}/a_\mathrm{ad} \to 0$ as $a\to a^\ast$, the system consisting of a dimer and $N-1$ bosons is expected to obey the same universal low-energy few-body physics around the atom-dimer resonance as that of an impurity and $N$ bosons near unitarity.
We will address this important point in Section~\ref{sec:Nbody}.

\section{Four-body problem}
\label{sec:4body}

In this section, we turn our attention to the four-body problem. Previous works have revealed the presence of tetramers tied to the ground-state Efimov trimer and have argued their universality.
For four identical bosons, it was found, both theoretically~\cite{hammer2007,Stecher2009,deltuva2011} and experimentally~\cite{Ferlaino2009,Pollack2009}, that there are two lower-lying tetramers tied to Efimov trimers.
There are also studies on the four-body problem where one atom of mass $M$ is resonantly interacting with three identical bosons of mass $m$.
For $M/m\le 1$, tetramer states have also been predicted to exist~\cite{Schmickler2017}. 
Importantly, when $M/m=1$, the tetramer spectrum is shown to be universal~\cite{Yoshida2018}; if we take the three-body parameter $a_-$ as the unit of length, different models have nearly equal tetramer energies as a function of the scattering length.
The opposite regime of the mass ratio $M/m>1$ has attracted much less attention.
Even though the Efimov effect occurs for arbitrary finite $M/m>1$~\cite{Braaten2006}, the existence of tetramers associated with the ground-state trimer has not been addressed.
This is partly because the bound states in this regime are too shallow to be captured numerically.
Here, we show in the limit of $M/m\to\infty$ that there is a tetramer state in a finite range of the inverse scattering length.
We also derive the asymptotic expression for its energy and argue its universality.

To solve the four-body problem,  we consider a
general four-body state that is written in the basis of $B_\k^\dag$:
\begin{align}
    |\Psi\rangle={\sum_{\k,\p,\q}}'\psi_{\k\p\q}B_{\k}^\dag B_{\p}^\dag B_{\q}^\dag|0\rangle,
\end{align}
where $\psi_{\k\p\q}$ is symmetric under permutation of the indices $\k$, $\p$ and $\q$. Substituting $|\Psi\rangle$ into the Schr\"{o}dinger equation, we obtain,
\begin{align}
    (E&-\epsilon_\k-\epsilon_\p-\epsilon_\q)\psi_{\k\p\q}=\label{4-body eq.}\\
    &U{\sum_{\u,\v}}'\left(\chi_{\k\p}^*\chi_{\u\v}\psi_{\u\v\q}+\chi_{\p\q}^*\chi_{\u\v}\psi_{\u\v\k}+\chi_{\q\k}^*\chi_{\u\v}\psi_{\u\v\p}\right).\nonumber
\end{align}

Similarly to the three-body problem, we may utilize the fact that the interaction term is a separable potential and rewrite the above equation,
\begin{eqnarray}
\psi_{\k\p\q}=\frac{\chi^*_{\k\p}f_\q+\chi^*_{\p\q}f_\k+\chi^*_{\q\k}f_{\p}}{E-\epsilon_\k-\epsilon_\p-\epsilon_\q},\label{4-body eq2.}
\end{eqnarray}
where $f_\q\equiv U{\sum_{\u\v}}'\chi_{\u\v}\psi_{\u\v\q}$. Substituting Eq.~\eqref{4-body eq2.} into this definition, we obtain
\begin{align}
    - Z(E-\epsilon_\q) f_\q
    = 2{\sum_{\k,\p}}'\frac{\chi_{\k\p}\chi_{\p\q}^*}
        {E-\epsilon_\k-\epsilon_\p-\epsilon_\q}f_\k,\label{4-body eq3.}
\end{align}
where we have taken the limit $U\to\infty$.
This is essentially a one-dimensional integral equation because $\chi_{\k\p}$ is independent of the direction of the momenta --- see Eqs.~\eqref{eq:eta} and~\eqref{eq:chi}.
Because the momentum summation ${\sum}'$ runs over $\mathbb{R}^3\cup\{i\kappa_\text{B}\}$, the integration kernel features 
a pole when $E>-2E_\text{B}$. We outline the details of how we deal with the pole in Appendix~\ref{ap:pole}.

\begin{figure}
    \centering
    \includegraphics[width=0.9\columnwidth]{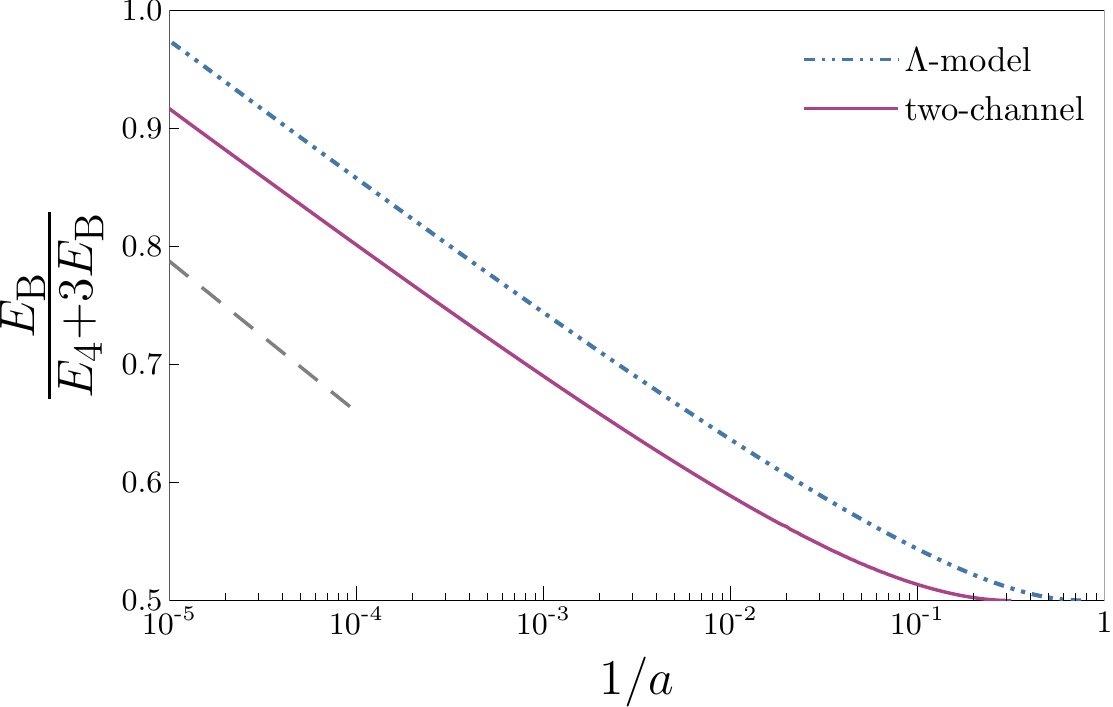}
    \caption{Inverse energy correction $\frac{E_\text{B}}{E_4+3E_\text{B}}$ as a function of $1/a$ calculated within the two-channel model (red solid) and the $\Lambda$~model (blue dash-dotted). The scattering length $a$ is in units of either $|r_0|$ or $1/\Lambda$ for the different models. The dashed line is a guide to the eye with slope $1/6\pi$.}
    \label{fig-tetramer}
\end{figure}

Remarkably, we may again derive an asymptotic form of the tetramer energy $E_4$ in the limit of $|r_0|/a\to 0$.
We first note that, in this limit, the wave function is concentrated at $\k=\p=\q=i\kappa_\mathrm{B}$.
Consequently, the amplitude $f_\q$ is negligible if $\q\in\mathbb{R}^3$, leading to the asymptotic energy equation,
\begin{align}
    -Z(E_4+E_\mathrm{B}) 
    \simeq {\sum_{\p}}'
        \frac{2 |\chi_{\p i\kappa_\mathrm{B}}|^2}
        {E_4+2 E_\mathrm{B}-\epsilon_\p} \!
    \simeq \frac{2 r_0^2/a^2}{E_4+3E_\mathrm{B}}.
\end{align}
Here we have dropped terms with $\p\in\mathbb{R}^3$ in the summation because they are negligible, as can be seen from a direct evaluation of the integral.
This together with Eq.~\eqref{eq:Z-asymp} yields 
\begin{align}
    E_4 \simeq -3 E_\mathrm{B} + \frac{6\pi E_\mathrm{B}}{\log(a/|r_0|)}.
    \label{eq:4Basymp}
\end{align}

Similarly to the trimer problem discussed in Sec.~\ref{sec:3body}, we see that on approaching unitarity, the tetramer energy goes to that of three non-interacting bosons interacting with a static potential. However, there is a very strong logarithmic correction due to the effective impurity-induced boson repulsion.
In Fig.~\ref{fig-tetramer} we show the resulting tetramer energy as a function of inverse scattering length. Again we plot a rescaled form $\eb/(E_4+3\eb)$, which $\sim \tfrac{\log a}{6\pi}$ according to Eq.~\eqref{eq:4Basymp}, and we show the results also within the $\Lambda$-model (Sec.~\ref{sec:lambda} and App.~\ref{ap:lambda}). Our results clearly show the model-independence of the logarithmic correction also in the four-body problem.

As seen in Fig.~\ref{fig-tetramer}, the tetramer exists only in a finite range of scattering lengths. Remarkably, this range is exactly the same as for the trimer, i.e., the tetramer merges into the atom-dimer continuum precisely at $a^*\simeq|r_0|/0.31821$ ($a^* \simeq 1.32/\Lambda$ in the $\Lambda$-model). 
This follows from the fact that the low-energy physics in the unitarity limit can be mapped to that at $a^*$~\cite{shortpaper}. 
As a result, while $1/a\to0^+$ corresponds to the point where the dimer, trimer, and tetramer unbind into free atoms, $a\to a^{*+}$ marks the point where the trimer and tetramer unbind into free atoms and a dimer state. We now generalize these results to the $(N+1)$-body problem.

\section{Asymptotic energy of an $(N+1)$-body bound state}
\label{sec:Nbody}

We can derive the asymptotic expression of the energy of an $(N+1)$-body bound state near unitarity by following a similar line of argument.
Starting from the \sch equation for the $(N+1)$-body system, we find an integral equation, which implicitly determines the energy $E_{N+1}$ of the $(N+1)$-body cluster:
\begin{align}
    &-Z(E_{\q_1\dots\q_{N-2}}) f_{\q_1\dots\q_{N-2}} \nn \\
    &=  {\sum_{\q_1',\q_2'}}' 
        \frac{\chi_{\q_1\q_2}\chi_{\q_1'\q_2'}f_{\q_1'\q_2'\q_3\dots\q_{N-2}}}
        {E_{\q_1\dots\q_{N-2}} - \epsilon_{\q_1'} - \epsilon_{\q_2'}}+
        (\q_1,\q_2)\stackrel{\substack{i<j\\2<j}}{\longleftrightarrow}(\q_i,\q_j) \nn \\
    &\quad + 2{\sum_{\q_1',\q_2'}}' 
        \frac{\chi_{\q_1\q_2'} \chi_{\q_1'\q_2'}^* f_{\q_1'\q_2\dots\q_{N-2}}}
        {E_{\q_1\dots\q_{N-2}} - \epsilon_{\q_1'} - \epsilon_{\q_2'}}
        + \q_1\stackrel{1<i}{\longleftrightarrow}\q_i,
        \label{eq:nbody}
\end{align}
where the symmetrization at the end of the second line results in ${N-2\choose2}-1$ terms, and that at the end of the third line in $N-3$ extra terms. Here, $n\choose r$ is the binomial coefficient and we have defined
$E_{\q_1\dots\q_{N-2}}\equiv E_{N+1}-\sum_{i=1}^{N-2} \epsilon_{\q_i}$,
Equation~\eqref{eq:nbody} generalizes Eq.~\eqref{eq:trimereng} for the trimer energy and Eq.~\eqref{4-body eq3.} for the tetramer energy to an arbitrary $N$.

For a large $N$, it becomes exponentially hard to solve Eq.~\eqref{eq:nbody} numerically to find the bound state energy. 
However, by noting that $f_{\q_1\dots\q_{N-2}}$ 
peaks around $\q_j=i\kappa_\mathrm{B}$ as $|r_0|/a\to0$, we can discard the terms in which this function has an argument $\q_i$ in $\mathbb{R}^3$.
This procedure yields 
\begin{align}
    -Z\left( E_{N+1}-(N-2)E_\mathrm{B} \right)
    \simeq \frac{(N-2)(N+1)r_0^2}{2a^2(E_{N+1}-NE_\mathrm{B})},
\end{align}
which, together with Eq.~\eqref{eq:Z-asymp}, leads to
\begin{align}
    E_{N+1}\simeq
    \left[
        - N
        + \binomial{N}{2} \frac{2\pi}{\log a}
    \right]E_\text{B},
    \quad 1/a\rightarrow 0^+,\label{asy_1}
\end{align}
where we have dropped the unit of length in $\log a$ since it is non-universal.
This expression agrees with our results for $N=2$ and $3$.
Furthermore, the binomial coefficient in front of the first correction
coincides with the number of possible trios composed of the impurity and two of the bosons in the $(N+1)$-body system.
Therefore, we conclude that the logarithmic correction originates from an effective three-body interaction.

Close to the atom-dimer resonance at $a^*$, we may also extract the asymptotic behavior~\cite{shortpaper} to find
\begin{align}
    E_{N+1}& \simeq-E_\text{B}+\left(-(N-1)+{N-1\choose 2}\frac{2\pi}{\log a_{\text{ad}}}\right)E_\text{B}^{\text{(ad)}},
    \label{asy_2}
\end{align}
when $a\to a^{*+}$.
Here $\eb^\text{(ad)}\equiv|E_3|-\eb$ is the trimer binding energy relative to the atom-dimer continuum. We see that this expression exactly mirrors Eq.~\eqref{asy_1}, which is a direct result of the equivalence of the low-energy physics at unitarity and at the atom-dimer resonance~\cite{shortpaper}.

Importantly, Eqs.~\eqref{asy_1} and \eqref{asy_2} indicate that there exist bound states below the dimer for any $N$ in the limit when $1/a\to0^+$ or when $a\to a^{*+}$. Furthermore, we have argued that the low-energy few-body physics is equivalent in the vicinity of these two scattering lengths. Given the special nature of both of these points, we have termed these \textit{multi-body resonances}~\cite{shortpaper}.
While for $N=2$ and 3 we have found precisely one bound state both within the two-channel model and the $\Lambda$-model, we remark that there could in principle be multiple bound states for larger $N$. Similarly, we have explicitly shown that the trimer and tetramer exist in the entire range $0<1/a<1/a^*$; while we cannot guarantee that this feature generalizes to larger $N$, it would appear plausible that this is the case.

\section{Conclusions and outlook}
\label{sec:conc}

To conclude, we have investigated the few-body problem of an
infinitely heavy impurity interacting with $N$ bosons. We have
demonstrated the existence of a bound state for \textit{any} $N$ when
the scattering length approaches unitarity, i.e., when $1/a\to0^+$,
and we have found that the inclusion of an effective three-body
repulsion generically leads to a strong logarithmic correction to the
binding energy. Specializing to the two-channel model, which is a standard tool
used to describe Fano-Feshbach resonances in ultracold gases, we have
furthermore found the existence of a second critical scattering
length $a^*$ beyond which all $(N+1)$-body bound states unbind into a
dimer and $N-1$ free bosons. We have argued that this can be
understood from
a formal equivalence between the problem of an infinitely
heavy impurity atom around
unitarity and the problem of an infinitely heavy dimer close to
$a^*$~(see also Ref.~\cite{shortpaper}). Moreover, this is a universal feature of any model that possesses an effective boson-boson repulsion and an atom-dimer resonance. In the special case of $N=2$, we have
demonstrated that the three-body problem may be solved analytically
both for the bound state and for scattering states,
and for $N=3$ we have calculated the tetramer energy exactly numerically.
By also calculating the three- and four-body spectrum within a second
model, the $\Lambda$-model, we have  demonstrated that the
qualitative features of our results are \textit{universal}, i.e.,
independent of the precise manner in which we introduce an effective
three-body repulsion.

One outstanding question of particular experimental relevance is to what extent
the physics described in this manuscript is dependent on the mass
ratio truly being infinite.
Here we note that while our results --- e.g.,~for the
asymptotic behavior of the bound state energies--- are exact in the limit
$M/m\to\infty$, some of the features such as the multi-body resonance
at the atom-dimer resonance $a^*$ may exist even when the mass ratio
is of order~1. For instance, by extending the calculation of
Ref.~\cite{Yoshida2018} we find that the locations of the atom-dimer
and atom-trimer resonances are numerically
indistinguishable both within the two-channel model and the 
$\Lambda$-model already when $M=m$~\cite{Yoshida2018}. Similarly, already for
equal masses, the ratio $|a_-|/a^*\simeq 2000$, implying that the
energy scale of Efimov trimers close to unitarity is exceedingly small in
typical experiments. Already a small change in mass ratio, for
instance to the experimentally relevant case of $^{40}$K impurities in
a $^{23}$Na gas, leads to an increase in this ratio by almost two
orders of magnitude. Therefore, with increasing mass ratio the three-body energy spectrum rapidly becomes reminiscent of our exact spectrum at $M/m=\infty$ in the sense that the trimers are only extremely weakly bound when $a<0$. However, we caution that it is not clear how far the analogy is valid, since it is not currently known
whether bound states even exist beyond the tetramer when $M=m$.

It is important to note that the length (and associated energy) scale that governs our results
(e.g., the logarithmic correction to the energy close to unitarity) can potentially be varied by orders of magnitude by looking at different Fano-Feshbach resonances. For a broad resonance, the typical length scale would be on the order of the van der Waals range, whereas for a narrow resonance this can take much larger values~\cite{Chin2010}.
This potentially allows the universal behavior
at the multi-body resonance points to
be probed at different Fano-Feshbach resonances. 
Indeed one important conclusion to draw from this work, is that the standard single-channel model dramatically fails to capture the physics of the impurity problem even exponentially close to unitarity.

Our predictions for the 
ground state energy can, for instance, be verified in cold-atom experiments by using radio-frequency pulses to
transfer the heavy impurity atom from a hyperfine state that is non-interacting with the surrounding bosons to an interacting state, and measuring the interaction energy. Indeed, such precision radio-frequency spectroscopy on an
impurity in a few-atom system
has already been carried out in one-dimensional
microtraps~\cite{Wenz2013}, and similarly there have been
precision measurements of multi-body interaction energies in optical lattice sites~\cite{Will2010,Mark2011,Goban2018}. 
In particular, it should be straightforward to verify that the ground-state energy
does not scale linearly with the number of majority bosons, which
would directly demonstrate the breakdown of the single-channel model for an infinitely heavy impurity.

\begin{acknowledgments}
We gratefully acknowledge fruitful discussions with Xiaoling Cui, Eugene Demler, Shimpei Endo, Victor Gurarie and Hui Zhai.
SMY acknowledges support from  the Japan Society for the Promotion of Science 
through Program for Leading Graduate Schools (ALPS) and 
Grant-in-Aid for JSPS Fellows (KAKENHI Grant No.~JP16J06706).
JL, ZYS, and MMP acknowledge financial support
from the Australian Research Council via Discovery Project
No.~DP160102739. JL is supported through the Australian
Research Council Future Fellowship FT160100244. 
JL and MMP acknowledge funding from the Universities Australia --
Germany Joint Research Co-operation Scheme.
\end{acknowledgments}

\appendix

\section{One-dimensional Anderson model with $p$-wave coupling\label{ap:p-wave}}

The coupling between $b_\k$ and $d$ is spherically symmetric in the bosonic Anderson model, $H_\mathrm{A}$ of Eq.~\eqref{eq:anderson}, which implies that we only need to retain the $s$-wave component and discuss the radial motion of the bosons.
This enables us to transform the bosonic Anderson model into an effective one-dimensional model.
More precisely, we can expand $b_\k$ in terms of the spherical harmonics,
\begin{align}
    b_\k 
    = \sum_{l=0}^\infty \sum_{m=-l}^l
        \frac{2\pi}{k} Y_l^m (\theta, \phi) b_{klm},
\end{align}
and substitute it into Eq.~\eqref{eq:single} to obtain
\begin{align}
    H_\mathrm{A0}
    &= \sum_{l=0}^\infty \sum_{m=-l}^l \int_0^\infty \frac{dk}{2\pi} \,
        \epsilon_k b_{klm}^\dag b_{klm}
    + \nu_0 d^\dag d \nn \\
    &\quad+ \frac{g}{\sqrt{2\pi^2}} \int_0^\infty \frac{dk}{2\pi} \, k \left(
        d^\dag b_{k00} + b_{k00}^\dag d
    \right).
    \label{eq:lm-decomp}
\end{align}
Here, $\theta$ and $\phi$ are the polar and azimuthal angles of $\k$, respectively, and $b_{klm}$ satisfies the following canonical commutation relations:
\begin{align}
    &[b_{klm}, b_{k'l'm'}^\dag] 
    = 2\pi\delta(k-k') \delta_{l,l'} \delta_{m,m'}, \\
    &[b_{klm}, b_{k'l'm'}] 
    = [b_{klm}^\dag, b_{k'l'm'}^\dag]
    = 0.
\end{align}
Equation~\eqref{eq:lm-decomp} makes it clear that only those bosons with $l=m=0$ couple to $d$. 
Thus we can drop all terms with $l>0$. 
Now, one can repeat a similar procedure of reduction to the radial motion, starting from the following one-dimensional Hamiltonian:
\begin{align}
    H_p
    &= \int_{-\infty}^\infty \frac{dk}{2\pi} \,
        \epsilon_k b_{k}^\dag b_{k}
    + \nu_0 d^\dag d \nn \\
    &\quad+ \frac{g}{2\pi} \int_{-\infty}^\infty \frac{dk}{2\pi} \, k\left(
        d^\dag b_{k} + b_{k}^\dag d
    \right)
    + \frac{U}{2} d^\dag d^\dag dd.
    \label{eq:pwave}
\end{align}
This one-dimensional model leads to exactly the same Hamiltonian for the radial motion in the $p$-wave sector as the one reached from Eq.~\eqref{eq:single} in the $s$-wave sector.
Therefore, the Anderson model~\eqref{eq:anderson} in three dimensions with $s$-wave coupling is equivalent to the one-dimensional model~\eqref{eq:pwave} with $p$-wave coupling.

\section{$\Lambda$-model}
\label{ap:lambda}

In the $\Lambda$-model~\cite{Yoshida2018}, an atom and an impurity interact via a single-channel zero-range potential or, equivalently, we can take $r_0\to 0$ with $a$ fixed in the two-channel model. 
The scattering length thus suffices to describe the two-body physics such as the dimer energy and the scattering amplitude.
However, in the three-body sector and beyond, regularization of a three-body interaction is necessary to obtain meaningful results if the mass of the impurity is finite, i.e., in the presence of the Efimov effect.
Moreover, we show in the main text that even in the limit of an impurity of infinite mass, the three-body interaction drastically affects the few-body phenomena.
To regularize the three-body interaction, the $\Lambda$-model restricts the momentum at which processes involving a boson and a dimer in the few-body equations can take place.

This procedure is easiest to explain by introducing the integral equations for the few-body problem. 
To this end, we start from the two-channel model~\eqref{eq:2ch}, focusing on the three-body problem; 
the generalization to more particles is straightforward.
One can reduce the \sch equation~\eqref{eq:sch-2ch-3body} to an integral equation by removing $\psi_c(\k_1, \k_2)$ and using Eq.~\eqref{eq:renorm} to eliminate the cutoff-dependent parameters:
\begin{align}
	T(E-\epsilon_\k)^{-1} \psi_d(\k)
	= \sum_\q \frac{\psi_d(\q)}{E-\epsilon_\k-\epsilon_\q}.
\end{align}
Here, $T(E)$ is the $T$ matrix~\eqref{eq:tmat}, which is effectively the dimer propagator renormalized by the two-body interaction.
In the $\Lambda$-model, we take $r_0 \to 0$ in the $T$ matrix.
The right-hand side can be viewed as 
the exchange of the impurity between the two bosons, or alternatively as exchanging which of the two bosons interacts with the impurity.
The two-channel model allows atoms with arbitrary momentum to participate in this process, having no restriction on the momentum summation.
On the other hand, in the $\Lambda$-model, the summation is truncated at $\Lambda$. 
This yields the integral equation for the three-body problem within the $\Lambda$-model~\cite{Bedaque1999}:
\begin{align}
    T_{r_0=0}(E-\epsilon_\k)^{-1} \psi_d(\k)
    = \sum_{\q, |\q|<\Lambda} 
        \frac{\psi_d(\q)}{E-\epsilon_\k-\epsilon_\q},1305.3182
\end{align}
where $T_{r_0=0}(E)^{-1}=\frac{m}{2\pi}(a^{-1}-\sqrt{-2mE})$.

By numerical solving this equation and its generalization to more particles, we found that there also exists an atom-dimer resonance at the two-body scattering length $a=a^*\simeq1.32/\Lambda$. The trimer state only exists when $a>a^*$. Moreover, the tetramer also merges into the atom-dimer continuum precisely at this resonance point.
 
The final results depend on the cutoff $\Lambda$, which is not a physical scale.
A renormalization procedure would in principle enable us to eliminate it by introducing a $\Lambda$-dependent counter term, which in this case is a three-body interaction~\cite{Bedaque1999}.
One can then take the limit $\Lambda\to\infty$ with fixed three-body observables such as the trimer energy $E_3$, and calculate other quantities in a $\Lambda$-independent manner.
Therefore, the $\Lambda$-model is equivalent to the single-channel model equipped with a three-body interaction.
In this paper, however, we do not follow this procedure as we are mainly interested in using the results of the $\Lambda$-model to emphasize the universal nature of our results.

\section{Pole structure of Eq.~\eqref{4-body eq3.}\label{ap:pole}}

Physically, the tetramer energy $E_4$ should satisfy
\begin{align}
    -3E_\text{B}<E_4<\min\{E_3,-E_\text{B}\}.
\end{align}
One might notice that on the right-hand side of Eq.~\eqref{4-body eq3.}, the integration will go through a pole if $E_4\geq -2E_\text{B}$ (recall that ${\sum}'$ runs over $\mathbb{R}^3\cup\{i\kappa_\text{B}\}$). In this region, the usual numerical algorithms for solving integral equations become unstable because of the divergent kernel near the pole.

In order to resolve this problem, we need to analyze the pole structure of the integral kernel and separate the corresponding nodes from $f_\k$, which can be done by defining the following auxiliary functions,
\begin{align}
    h_\k\equiv\frac{|\chi_{i\kappa_\text{B}i\kappa_\text{B}}|^2}{E+2E_\text{B}-\epsilon_\k}\left(\frac{f_\k}{\eta_\q^*}+2\frac{f_{i\kappa_\text{B}}}{\eta_{i\kappa_\text{B}}^*}\right),\quad
    h_{i\kappa_\text{B}}\equiv\frac{f_{i\kappa_\text{B}}}{\eta_{i\kappa_\text{B}}^*}.\nonumber
\end{align}

Note that the coefficient in front of the bracket has the exact same pole structure as the integral kernel of Eq.~\eqref{4-body eq3.}. Thus it cancels the zero of $f_\k$ and transforms the integral equation into
\begin{widetext}\begin{align}
    \left[(2I_1(\epsilon_\q)+I_2(\epsilon_\q))(E+2E_\text{B}-\epsilon_\q)+|\chi_{i\kappa_\text{B}i\kappa_\text{B}}|^2\right]h_\q&=\nonumber\\
    \sum_{\k}\frac{-2|\eta_\k|^2}{|\eta_{i\kappa_\text{B}}|^2}(I_0(E_\text{B}+\epsilon_\q+\epsilon_\k)+I_1(E_\text{B}+&\epsilon_\q+\epsilon_\k))(E+2E_\text{B}-\epsilon_\k)h_\k+6|\chi_{i\kappa_\text{B}i\kappa_\text{B}}|^2(I_1(\epsilon_\q)+I_2(\epsilon_\q))h_{i\kappa_\text{B}},\nonumber\\
    3(I_2(-E_\text{B})-I_0(-E_\text{B}))h_{i\kappa_\text{B}}=\sum_{\k}&\frac{2|\eta_\k|^2}{|\eta_{i\kappa_\text{B}}|^2|\chi_{i\kappa_\text{B}\kappa_\text{B}}|^2}\left[I_1(\epsilon_\k)(E+2E_\text{B}-\epsilon_\k)+|\chi_{i\kappa_\text{B}i\kappa_{\text{B}}}|^2\right]h_\k.\label{eq.apc1}
\end{align}\end{widetext}
Here $I_i(\epsilon_\q),\  i=0,1,2$ are defined as
\begin{align}
    I_0(\epsilon_\q)&\equiv\frac{|\chi_{i\kappa_\text{B}i\kappa_\text{B}}|^2}{E+2E_\text{B}-\epsilon_\q},\\
    I_1(\epsilon_\q)&\equiv\sum_\p\frac{|\chi_{pi\kappa_\text{B}}|^2}{E+E_\text{B}-\epsilon_\p-\epsilon_\q},\\
    I_2(\epsilon_\q)&\equiv\sum_{\k,\p}\frac{|\chi_{\k\p}|^2}{E-\epsilon_\k-\epsilon_\p-\epsilon_\q},
\end{align}
which can all be calculated analytically in a manner similar to $Z(E)$.

Both Eq.~\eqref{eq.apc1} and Eq.~\eqref{4-body eq3.} give the correct tetramer energy. However, the former has no singularity in the region $E>-2E_\text{B}$, which makes it more suitable for numerical studies.

\bibliography{bosepolaron}

\end{document}